\documentclass[twocolumn]{aastex631}
\usepackage{amsmath}
\usepackage{booktabs}
\usepackage{mhchem}
\usepackage{gensymb}
\usepackage[T1]{fontenc}

\shorttitle{Characterizing L-type Asteroids}
\shortauthors{Gomez Barrientos et al.}

\begin{document}

\title{Detection of a 2.85 micrometer Feature on 5 Spinel-rich Asteroids from JWST}

\correspondingauthor{Jonathan Gomez Barrientos}
\email{jgomezba@caltech.edu}

\author[0000-0002-0672-9658]{Jonathan Gomez Barrientos}
\affiliation{Division of Geological and Planetary Sciences, California Institute of Technology, Pasadena, CA 91125, USA}

\author[0000-0002-9068-3428]{Katherine de Kleer}
\affiliation{Division of Geological and Planetary Sciences, California Institute of Technology, Pasadena, CA 91125, USA}

\author[0000-0002-2745-3240]{Bethany L. Ehlmann}
\affiliation{Division of Geological and Planetary Sciences, California Institute of Technology, Pasadena, CA 91125, USA}

\author[0000-0001-6622-2907]{Francois L.H. Tissot}
\affiliation{The Isotoparium, Division of Geological and Planetary Sciences, California Institute of Technology, Pasadena, CA 91125, USA}

\author{Jessica Mueller}
\affiliation{Division of Geological and Planetary Sciences, California Institute of Technology, Pasadena, CA 91125, USA}

\begin{abstract}

Ground-based observations of `Barbarian' L-type asteroids at 1 to 2.5-$\mu$m indicate that their near-infrared spectra are dominated by the mineral spinel, which has been attributed to a high abundance of calcium-aluminum inclusions (CAIs) --- the first solids to condense out of the protoplanetary disk during the formation of the Solar System. However, the spectral properties of these asteroids from 2.5 to 5-$\mu$m, a wavelength region that covers signatures of hydrated minerals, water, and organics, have not yet been explored. Here, we present 2 to 5-$\mu$m reflectance spectra of five spinel-rich asteroids obtained with the NIRSpec instrument on the James Webb Space Telescope. All five targets exhibit a $\sim$ 2.85-$\mu$m absorption feature with a band depth of 3-6\% that appears correlated in strength with that of the 2-$\mu$m spinel absorption feature. The shape and position of the 2.85-$\mu$m feature are not a good match to the 2.7-$\mu$m feature commonly seen in carbonaceous CM meteorites or C-type asteroids. The closest spectral matches are to the Moon and Vesta, suggesting commonalities in aqueous alteration across silicate bodies, infall of hydrated material, and/or space weathering by solar wind H implantation. Lab spectra of CO/CV chondrites, CAIs, as well as the minerals cronstedtite and spinel, also show a similar feature, providing clues into the origin of the 2.85-$\mu$m feature.

\end{abstract}


\section{Introduction}
\label{sec:intro}

The Solar System’s small body populations hold a record of the chemical and dynamical processes at play during the first few million years of its history. The stony S-class asteroids have surfaces composed of silicate minerals; their near-infrared spectra are typically dominated by olivine and pyroxene, but a small subset of the S-complex known as the L class has a near-infrared spectrum that is dominated by the mineral spinel. The L asteroid class was introduced by \cite{bus2002} and is distinguished in reflectance data by a steep optical slope short of 0.75-$\mu$m, the lack of a deep 1-$\mu$m absorption feature typically seen on S-types, and sometimes a shallow 2-$\mu$m absorption feature \citep{DeMeo2009}. The ratio of the 1 and 2-$\mu$m band depths is not consistent with the spectral ratios expected for the olivine/pyroxene mixtures present on typical S-class asteroids \citep{burbine1992}, and is instead attributed to Fe in spinel \citep{burbine1992, Sunshine2008}. Many of the objects with this spectral property also exhibit anomalous visible and near-infrared polarization curves, which are consistent with the wavelength-dependent index of refraction of the mineral spinel \citep[e.g.][]{cellino2006,Devogele2018,masiero2023}. Asteroids that exhibit this polarization signature are designated ‘Barbarians’ for the asteroid 234 Barbara. Moving forward, we adopt the terminology `spinel-rich' asteroids to refer to objects that are classified as L-class and contain the ‘Barbarian’ polarization signature.

A high abundance of spinel may be diagnostic of a parent body magmatic process \citep[e.g.,][]{Pieters2011, Gross2011}, or alternatively may indicate the presence of calcium aluminum inclusions (CAIs), in which spinel is a primary constituent \citep{Sunshine2008}. Spinel can have a large impact on the reflectance spectrum of a surface even in small abundance because of its high reflectivity, and under the CAI explanation for the spinel, \cite{burbine1992} estimated a surface abundance of 5-10\% of CAI's for their two targets. \cite{Sunshine2008} compared additional asteroid spectra with models based on laboratory spectra of CAIs from the meteorite Allende; they estimated 3 asteroid families with 30$\pm$10\% CAI surface compositions. \cite{Devogele2018} presented additional 1 to 2.5-$\mu$m spectra and modeled these along with previously obtained spectra of L-type asteroids by the SMASS \citep{Xu1995, bus2002} and MIT-Hawaii-IRTF NEO surveys \citep{Binzel2006}. Using mixtures of lab spectra, \cite{Devogele2018} found that their surfaces were also consistent with tens of percent of CAIs. These surface abundances are higher than the $<5$\% abundance in known meteorites \citep{Hezel2008, Rubin2011}, and would imply specific nebular conditions during these objects’ formation and revise understanding of accretion processes during a very early phase of the Solar System's formation history.

Compared to 0.4 to 2.5-$\mu$m (i.e., basis of \citealt{bus2002, DeMeo2009} classifications), the 2.5 to 5-$\mu$m wavelength range also encodes crucial but separate information about the composition and processes that affect asteroids, reflecting the environments in which they formed. In particular, the 2.5 to 5-$\mu$m range records absorptions from OH, water, organics, carbonate, and other volatiles. \ce{H2O} exhibits a broad, shallow feature centered around 3.1-$\mu$m, while metal-OH, including in hydrated minerals that formed via aqueous alteration, exhibit a sharper, asymmetric feature centered around 2.7-$\mu$m \citep[e.g.,][]{Clark1999}. The 3-$\mu$m region has been studied extensively on the Moon, which shows an absorption feature near 2.85-$\mu$m that has been linked solar wind implantation or surficial \ce{H2O} \citep{Sunshine2009, Pieters2009, Clark2009, McCord2011, Wohler2017, Honniball2020, Chauhan2021, Laferriere2022, Wilk2024}. The 3-$\mu$m band has also been extensively studied on the volatile-rich, carbonaceous C-type asteroids, some of which exhibit absorption at 2.7-$\mu$m or 3.1-$\mu$m with band depths as high as tens of percent \citep[e.g.,][]{Rivkin2010, Campins2010, Takir2012, Kitazato2019, Rivkin2022}. Despite the strong 3-$\mu$m features of the C-types, the obstruction by Earth’s atmosphere across 2.5 to 2.8-$\mu$m has challenged the interpretation of the OH mineral feature in particular, since the band minimum and the slope of the short-wavelength side of the band are obscured. 

The James Webb Space Telescope (JWST), with its high sensitivity and access to wavelengths typically blocked by Earth's atmosphere, enables the detection and characterization of shallower bands and across a wider range of asteroid classes than previously possible. Here, we used the NIRSpec instrument \citep{Jakobsen2022} on JWST to observe the 2.5 to 5-$\mu$m region of five spinel-rich asteroids, in order to characterize their geologic histories and provide new constraints on their surface compositions.

Our observations and data reduction procedures are described in Section \ref{sec:data}. The results are briefly presented in Section \ref{sec:res}, and discussed in Section \ref{sec:disc}, including comparisons to hydration features observed on meteorites and other classes of asteroids. The conclusions are summarized in Section \ref{sec:conc}.

\section{Data Collection and Analysis}
\label{sec:data}

\subsection{JWST Observations}

We collected 1.7 to 5.1-$\mu$m spectra of five L-type asteroids as part of the JWST Cycle 1 GO 2361 program to look for evidence of water, hydroxyl, or other volatiles on their surfaces. The target list consists of asteroids that span 10-40 km in diameter and are located between 2.7-3.1 AU. These targets (except for 1040 Klumpkea) have been previously  observed spectroscopically from the ground. In particular, the near-infrared spectra of 458 Hercynia and 1372 Haremari were taken by \cite{Devogele2018} while the near-infrared spectra of 824 Anastasia and 2085 Henan were taken by the MIT-Hawaii-IRTF NEO survey \citep{Binzel2006}. The visible spectra of all four objects were acquired by the SMASS surveys \citep{Xu1995, bus2002}. Furthermore, all five targets exhibit a large polarization inversion angle or belong to asteroid families that exhibit this signature.

The JWST observations of these five L-type asteroid were acquired using NIRSpec on JWST in fixed-slit mode (S200A1). We used the gratings G395M (R$\sim$1000) and G235M (R$\sim$1000) combined with the filters F290LP (2.9 to 5.1-$\mu$m) and F170LP (1.7 to 3.1-$\mu$m). In Table \ref{tab:obs} we present the target list along with the observation details.

\subsection{JWST Data Reduction}

We extract calibrated 1D spectra from all targets in our survey using the JWST Science Calibration Pipeline Version 1.14.1.dev8 \citep{Howard_2023}. The pipeline initially applies standard detector-level corrections (e.g., linearity) to the raw 2D images to obtain uncalibrated count rate images (stage 1). The count rate images contain 1/f noise \citep[e.g.,][]{Rauscher2012}, which is not removed by the pipeline. We remove this noise component by first taking a small section at the bottom of the count rate image and calculating the median of each of its columns. We then subtract the median of each column from the entire count rate image. With the 1/f noise removed, the pipeline then applies physical corrections (e.g., background subtraction) to produce calibrated count rate images (stage 2). In the final stage, the pipeline produces rectified 2D spectra, from which it extracts the corresponding 1D spectra. Given our observational strategy, for each target, the pipeline extracted 1D spectra separately for each filter (F170LP and F290LP) and dither position (3 or 5 point-nod) on the detector. 

\begin{figure}
    \centering
    \includegraphics[width=0.95\columnwidth]{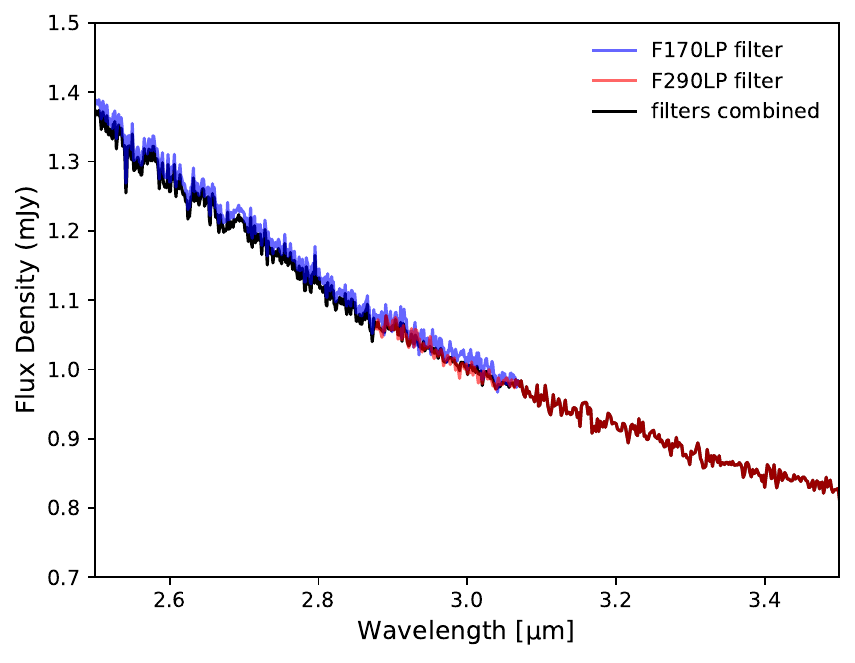}
    \caption{Processing of the two NIRSpec spectra of 824 Anastasia from the filters F170LP and F290LP. A small offset is applied to the entire F170LP data in order to match the F290LP data in the overlapping region.}
    \label{fig:join}
\end{figure}

Having obtained 1D spectra, we designed a custom pipeline to produce a combined 1D spectrum for each target that spans the combined wavelength range of the F170LP and F290LP filters. Our pipeline first identifies outliers in each dithered spectrum by using a moving median filter of 21 pixels, large enough to identify outliers. Data points that exceed 3 standard deviations from the median of the filter are flagged and masked. These outliers are then replaced by the median value of neighboring pixels. For a given filter, we then combine all of the dithered spectra (3 or 5 depending on the target) into a single spectrum by taking their mean. In cases where there is an offset between a spectrum and the other dithered spectra, we calculate the offset and remove it from the offset spectrum before calculating the mean spectrum for a given filter. Then, for each target, we scale the F170LP data to the F290LP data (typically 1-3 $\%$ depending on the target) to obtain a continuous spectrum that spans the combined wavelength coverages of the two filters as shown in Figure \ref{fig:join}). In the region where the wavelength coverage of the F170LP and F290LP overlap (2.88 to 3.07-$\mu$m), we interpolate the spectrum with higher spectral resolution to the wavelength grid of the spectrum with lower resolution and take the mean flux value between the two datasets.

 \subsection{Continuum Modeling}

We model the reflected and thermal continuum of the JWST spectra to produce a spectrum in normalized reflectance units for identification of spectral features. Reflected sunlight dominates the continuum at shorter wavelengths, while thermal emission from the asteroids dominates at longer wavelengths. The reflected sunlight (in units of Janskys; $\mathrm{10^{-26}\; W\cdot Hz^{-1} \cdot m^{-2}}$) at any given frequency (or wavelength) is

\begin{equation}
    \mathrm{Reflected \; Flux \; Density \,} = \frac{\mathrm{F_{\odot_\nu}}}{\mathrm{r^2}}\frac{\mathrm{R^2_{aster}A_g}}{\mathrm{R^2}}\mathrm{\Phi (\alpha)}
\end{equation}

where $\mathrm{F_{\odot_\nu}}$ is the frequency-dependent solar flux at 1AU (in Janskys), $\mathrm{r}$ is the distance between the Sun and the asteroid (in AU), $\mathrm{R_{aster}}$ is the volume equivalent radius of the asteroid, $\mathrm{A_g}$ is the geometric albedo of the asteroid, $\mathrm{R}$ is the distance between the observer and the asteroid (in the same units as $\mathrm{R_{aster}}$), and $\mathrm{\Phi}$ is a frequency-independent phase correction factor, which depends on the phase angle $\mathrm{\alpha}$. For $\mathrm{F_{\odot_\nu}}$, we adopt the stellar spectrum of P330 (Program 1538), a standard G2V star, observed with the same NIRSpec filter/grating combinations as our program. Ideally, the stellar spectrum would be divided from the target data at full spectral resolution to remove solar features in the reflectance spectra of our targets. However, in practice this increases the noise in the target spectra. We therefore smooth the stellar spectrum by 21 pixels to reduce noise while also preserving the spectral shape before dividing our target spectra by it, but show the full-resolution stellar spectrum alongside the data when we present results so that the solar spectral features in the target spectra can be easily identified. After smoothing, the stellar spectrum is scaled to a flux of $3.29\times 10^{16}$ mJy at 4.85$\mu$m (\citealt{Tokunaga2000}, \citealt{Cox2000}). For each object, we obtained $\mathrm{R}$, $\mathrm{r}$, and $\mathrm{\alpha}$ at the time of observation from the Jet Propulsion Laboratory Horizons online ephemeris generator. Finally, we adopted Equation 21 of \citealt{Bowell1989} as the phase correction factor. This equation consists of parameters (w, $\mathrm{B_0}$, h, b, and $\mathrm{\bar{\theta}}$) that depend on the spectral class of an asteroid. There are no L-type asteroids for which these parameters are reported (see Table IV of \citealt{Bowell1989}), but since S-type asteroids are the closest taxonomical neighbors to the L-types, we selected parameter values of the S-type asteroid 133 Cyrene. As a result, we calculated $\mathrm{\Phi}$ values in the range of 0.36 and 0.44 for our targets.

We model the thermal component of the spectra with the Near Earth Asteroid Thermal Model (NEATM; \citealt{Harris1998}). The output of the NEATM is a temperature profile $\mathrm{T=T_{max}\cos^{1/4} \theta_i}$, where $\theta_i$ is the incidence angle. The maximum temperature (in Kelvin) is equal to

\begin{equation}
    \mathrm{T_{max}} = \frac{(1-\mathrm{A_B})\mathrm{S_\odot}}{\mathrm{\eta \sigma \epsilon r^2}}
\end{equation}

where $\mathrm{S_\odot}$ is the solar constant at 1AU (1367.2 $\mathrm{W/m^2}$) and $\mathrm{\sigma}$ is the Stefan-Boltzmann constant. The beaming parameter $\mathrm{\eta}$ largely encapsulates the observed increase in thermal emission from asteroids at small phase angles due to their rough surfaces \citep{Hansen1977, Lebofsky1986}. The bolometric bond albedo $\mathrm{A_B}$, can be related to the geometric albedo of the reflected-light model via the relation $\mathrm{A_{B}} = (0.29+0.684*\mathrm{G})*\mathrm{A_g}$ \citep{Bowell1989}. We assume a default value of 0.15 for the slope parameter G and a constant spectral emissivity $\mathrm{\epsilon}$ of 0.9 for all objects. Finally, the thermal flux at each frequency (or wavelength) is determined by integrating the Planck function over the asteroid's illuminated hemisphere that lies in the field of view of the observer as dictated by the phase angle. The asteroid is modeled as a faceted sphere, with the temperature of each facet set by the NEATM model.

\begin{deluxetable*}{lcccccccccc}[t!]
\tablewidth{0pc}
\setlength{\tabcolsep}{5pt}
\tabletypesize{\normalsize}
\renewcommand{\arraystretch}{1.1}
\tablecaption{
    JWST Observation and Target Details
    \label{tab:obs}
}
\tablehead{ \colhead{Target} & \multicolumn{1}{c}{UT Date-time\tablenotemark{\scriptsize a}} & \multicolumn{1}{c}{$D$ (km)\tablenotemark{\scriptsize b}} & \multicolumn{1}{c}{$A_{g}$ \tablenotemark{\scriptsize b}} &
\multicolumn{1}{c}{$\eta$ \tablenotemark{\scriptsize b}} &  \multicolumn{1}{c}{$r$ (au)\tablenotemark{\scriptsize c}} & \multicolumn{1}{c}{$R$ (au)\tablenotemark{\scriptsize c}}& \multicolumn{1}{c}{$\alpha$ (deg)\tablenotemark{\scriptsize c}} 
    & \multicolumn{1}{c}{dithers\tablenotemark{\scriptsize c}} & \multicolumn{1}{c}{$t_{\mathrm{exp}}$ (s)\tablenotemark{\scriptsize d}} 
}
\startdata
458 Hercynia & 2023-08-31 07:00 & 25 & 0.5 & 0.60 & 3.51 & 3.30 &  16.8 & 5-point & 1278 \\
824 Anastasia & 2022-11-27 23:52 & 33 & 0.18 & 0.74 & 3.12 & 2.82 & 18.4 & 3-point & 1150 \\
1040 Klumpkea  & 2022-11-05 11:12 & 16 & 0.68 & 0.60 & 2.84 & 2.18 & 17.3 & 5-point & 2556 \\
1372 Haremari & 2022-10-28 15:45 & 22 & 0.18 & 0.89 & 2.70 & 2.53 & 21.8 & 5-point & 2556 \\
2085 Henan & 2023-01-10 11:47 & 13 & 0.18 & 0.86 & 2.47 & 2.29 & 23.7 & 5-point & 2556 \\
\enddata
\textbf{Notes.}
\vspace{-0.15cm}\tablenotetext{\textrm{a}}{The time corresponds to the observation mid-time (hour-minute).}
\vspace{-0.15cm}\tablenotetext{\textrm{b}}{Diameter $D$, infrared geometric albedo $A_{g}$, and beaming parameter $\eta$,values are derived from our thermal modeling.}
\vspace{-0.15cm}\tablenotetext{\textrm{c}}{Heliocentric distance $r$, distance from JWST $R$, phase angle $\alpha$, correspond to the time of observation.}
\vspace{-0.15cm}\tablenotetext{\textrm{d}}{Total exposure times across all dithers and grating settings.}
\vspace{-0.5cm}
\end{deluxetable*}
\begin{figure}
    \centering
    \includegraphics[width=0.95\columnwidth]{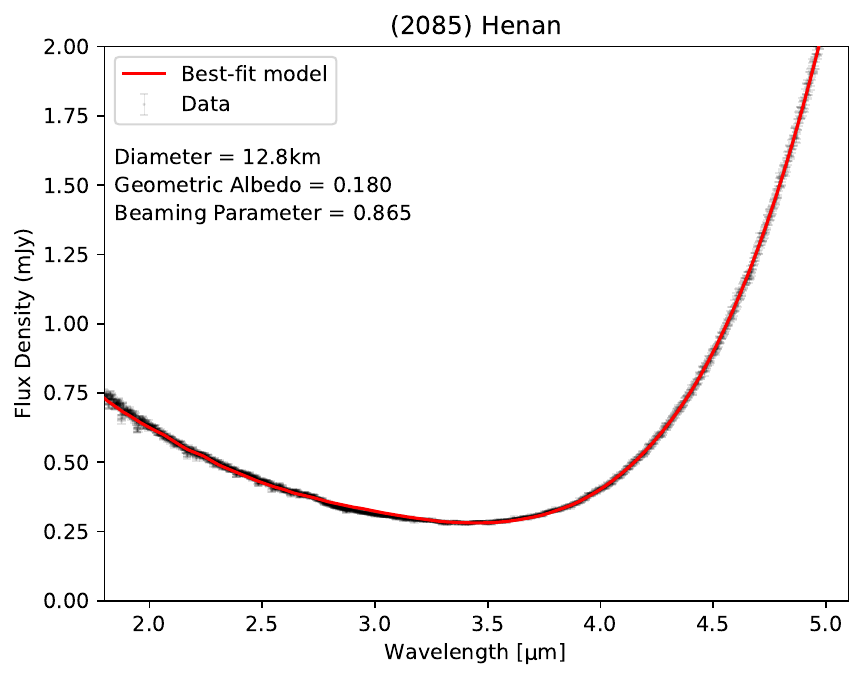}
    \caption{Modeling of the continuum of 2085 Henan. The black points correspond to the JWST data. The red curve is the best-fit continuum model, which consists of a reflected-light model and a thermal emission model.}
    \label{fig:best-fit}
\end{figure}
\vspace{-8mm}
Having constructed models for the continuum of the spectra, we fit the data simultaneously for the reflected-light and thermal emission. We fit the data by minimizing $\chi^2$, with $\mathrm{R_{aster}}$, $\mathrm{A_g}$, and $\eta$ as the free parameters. For fitting, we estimate the uncertainties on the datapoints by dividing each spectrum by a smoothed version of itself and then calculating the standard deviation of the residuals. We also estimated uncertainties using the JWST Science Calibration Pipeline but found that these uncertainties are largely underestimated.  In Figure \ref{fig:best-fit} we present the best-fit model of 2085 Henan and in Table \ref{tab:obs}, we list the best-fit parameter values of all five targets. We note that the best-fit diameters and albedos do not always agree with previous estimates \citep[e.g.,][]{Masiero2014}, perhaps due to changes in the viewing geometry of these irregular-shaped objects, variations in their phase function, and the absolute flux calibrations of the JWST observations. Finally, after fitting, we remove the thermal excess from the spectra by subtracting the best-fit thermal model from the data. We then normalize the data with respect to the best-fit reflected-light model (i.e., divide out the reflectance model).

\subsection{Measuring Band Properties}
\label{sec:fitting_feature}

We quantify the 3-$\mu$m band properties (band depth, area, and position) in the spectra as follows. We initially model and fit the data in the 3-$\mu$m band with the composite of a Gaussian and a linear component using the python package \texttt{LMfit}. Due to the asymmetric shape of the absorption features, we find that this composite model yields an inadequate fit. Therefore, we then construct a composite model that equals the sum of an asymmetric Gaussian and a linear component. The asymmetric Gaussian component is 

\begin{equation} \label{eq:3}
    F(\lambda) = \begin{cases} 
      A*\exp{\frac{-(\lambda - \mu)^2}{2\sigma_1^2}} & \lambda \leq \mu \\
      A*\exp{\frac{-(\lambda - \mu)^2}{2\sigma_2^2}} & \lambda > \mu 
   \end{cases}
\end{equation}

where \textit{A}, $\mu, \sigma_1,$ and $\sigma_2$ are free parameters. For the linear component, the slope and the intercept are the free parameters. We report the band center (\textit{BC}) as $\mu$ from the best-fit model and report the uncertainty from \texttt{LMfit}. We calculate the band depth as $1 - \frac{RB}{RC}$ where $RB$ is the reflectance predicted by the composite model (linear + asymmetric Gaussian components) at $\lambda=BC$ and $RC$ is the reflectance predicted by the linear model at $\lambda=BC$. We calculate the uncertainty in the band depth by first propagating the uncertainties in BC, the amplitude of the asymmetric Gaussian, the intercept, and slope of the linear model to calculate the uncertainty in $RB$ and $RC$. We then propagate these uncertainties through $1 - \frac{RB}{RC}$ to estimate the uncertainty in BD. Finally, we calculate the band area by integrating the linear component and the composite model and then subtracting the two areas.

\subsection{Laboratory Spectra}

To support the interpretation of the JWST data, we collected infrared spectra of two CAI specimens for comparison. The two CAIs originate from the CV3 chondrites Allende and Leoville \citep[e.g.,][]{Johnson1990}. In-situ mid-infrared reflectance measurements of CAI specimen Leoville ME2628-12 and Allende ME2629-4.167 were acquired using Caltech’s Nicolet iS50 micro-FTIR. An MCT-B detector and CsI beam splitter were used to measure a wavelength range from 2.5 to 25-$\mu$m with a resolution of 4 cm$^{-1}$. Background contributions were corrected for using a gold mirror and each spectrum is an average of 200 scans. The aperture size used was 50-$\mu$m. The surface of Leoville ME2628-12 was polished while the surface of Allende ME2629-4.167 was a rough slab. During data acquisition, a plastic cover enclosed samples to limit atmospheric contributions.

\section{Results}
\label{sec:res}

\begin{figure*}
    \centering
    \includegraphics[width=0.95\textwidth]{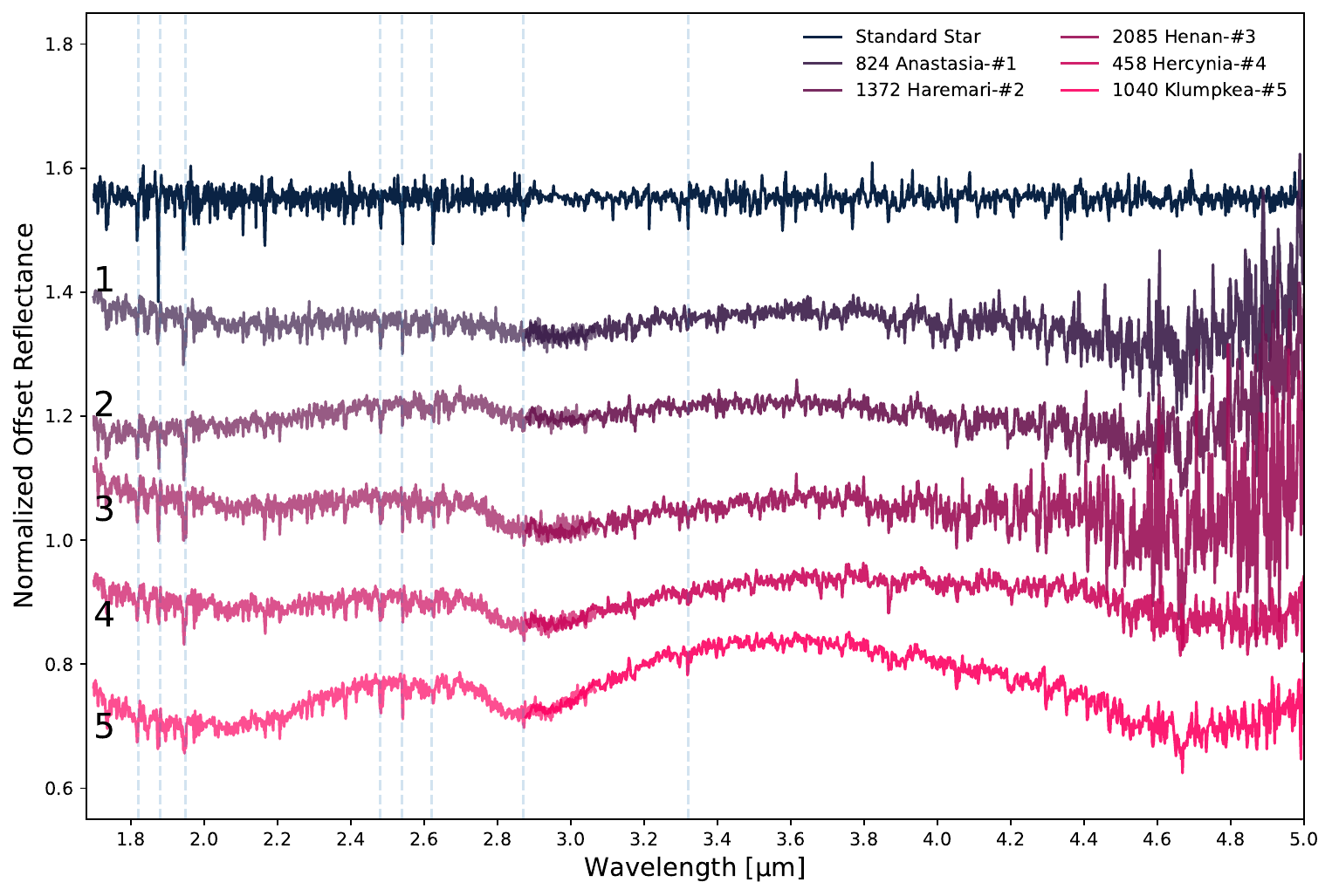}
    \caption{JWST NIRSpec spectra of five L-type asteroids. The spectra have been corrected for thermal emission and normalized with respect to a reflected sunlight model. The light and dark colored spectra correspond to the data acquired with the F170LP and F290LP filters, respectively. The five targets exhibit an absorption feature near 3-$\mu$m, which could be due to \ce{OH}/\ce{H2O} on their surfaces. The dashed blue lines indicate potential detector artifacts.}
    \label{fig:norm_reflec}
\end{figure*}

The normalized reflectance spectra of the five asteroids show an absorption feature near 3-$\mu$m (Figure \ref{fig:norm_reflec}). We measure the properties of this feature across the sample (see Section \ref{sec:fitting_feature}) and list them in Table \ref{tab:band_properties}. In general, the feature is sharp and asymmetric. The feature on all asteroids has a band center (minimum) of $\sim$2.85-$\mu$m, which is generally consistent with the presence of \ce{OH} and/or adsorbed \ce{H2O} on their surfaces. However, the strength of the 2.85-$\mu$m feature varies in the sample. The most pronounced feature is found in the spectrum of 1040 Klumpkea (6.75 $\%$ band depth) and the most subtle feature is found in the spectrum of 824 Anastasia (2.77 $\%$ band depth). Lastly, 824 Anastasia and 1372 Haremari also appear to exhibit spectral features between 3.2 and 3.6-$\mu$m, a spectral range that covers signatures of organics \citep[e.g.,][]{Kaplan2021}. However, we note that these features are not present across all dithered spectra, and we therefore consider them to be artifacts.

In addition to the 3-$\mu$m feature, the normalized reflectance spectra of 1040 Klumpkea, 458 Hercynia, and 2085 Henan show an absorption feature near 2-$\mu$m, which has been linked to spinel in CAIs. In our sample, the strength of the 2-$\mu$m feature is correlated with the strength of the 3-$\mu$m feature. Asteroid 1040 Klumpkea, which has the strongest 3-$\mu$m feature, also has the strongest 2-$\mu$m feature. Furthermore, 458 Hercynia and 2085 Henan, which have similar 3-$\mu$m band depths, also have similar 2-$\mu$m features. This is consistent with previous modeling of ground-based spectra indicating that 458 Hercynia and 2085 Henan contain similar amounts of CAIs \citep[see][]{Devogele2018}. Despite modeling by \cite{Devogele2018} also indicating that 824 Anastasia and 1372 Haremari contain CAIs, it is unclear if a 2-$\mu$m feature is present in their JWST spectra.

\begin{deluxetable}{llll}[t!]
	\tablecaption{The measured 3-$\mu$m band properties of five L-type asteroids.}
	\label{tab:band_properties}
	\tablewidth{0pt}
	\tablehead{\colhead{Target} & \colhead{Band Center} & \colhead{Band Depth} & \colhead{Band Area} \\
                    & ($\mu$m) & ($\%$) & ($\mu$m$^{-1}$)}
	\startdata
	458 Hercynia & 2.85$\pm{0.01}$ & 4.89$ \pm{0.05}$ & 0.019 \\
        824 Anastasia & 2.93$\pm{0.02}$ & 2.77$ \pm{0.02}$ & 0.012 \\
        1040 Klumpkea & 2.87$\pm{0.01}$ & 6.75$ \pm{0.10}$ & 0.024 \\
        1372 Haremari & 2.86$\pm{0.01}$ & 2.94$ \pm{0.05}$ & 0.012 \\
        2085 Henan & 2.86$\pm{0.01}$ & 4.68$ \pm{0.04}$ & 0.023 \\
	\enddata
\end{deluxetable}

\begin{figure}
    \centering
    \includegraphics[width=0.95\columnwidth]{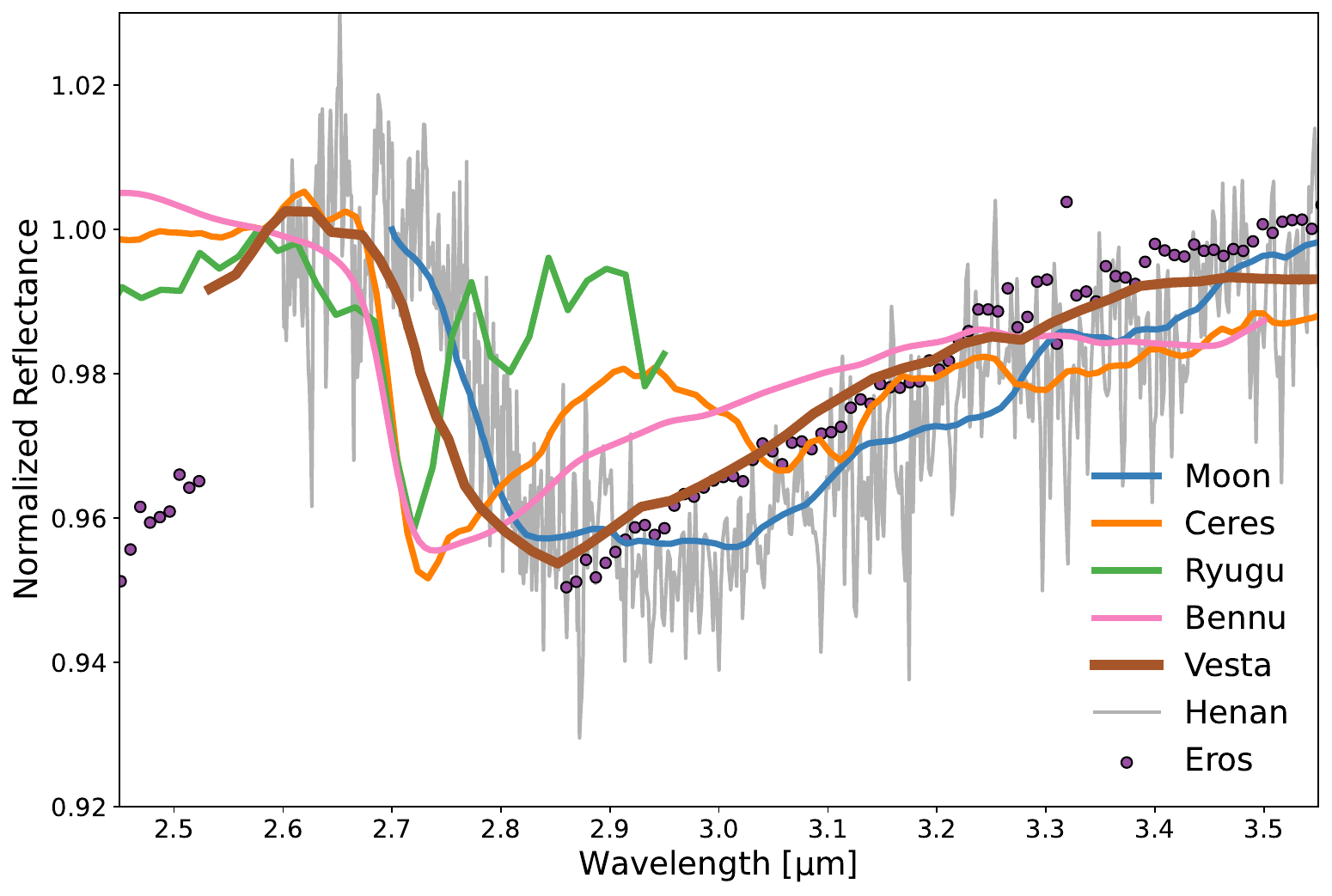}
    \includegraphics[width=0.95\columnwidth]{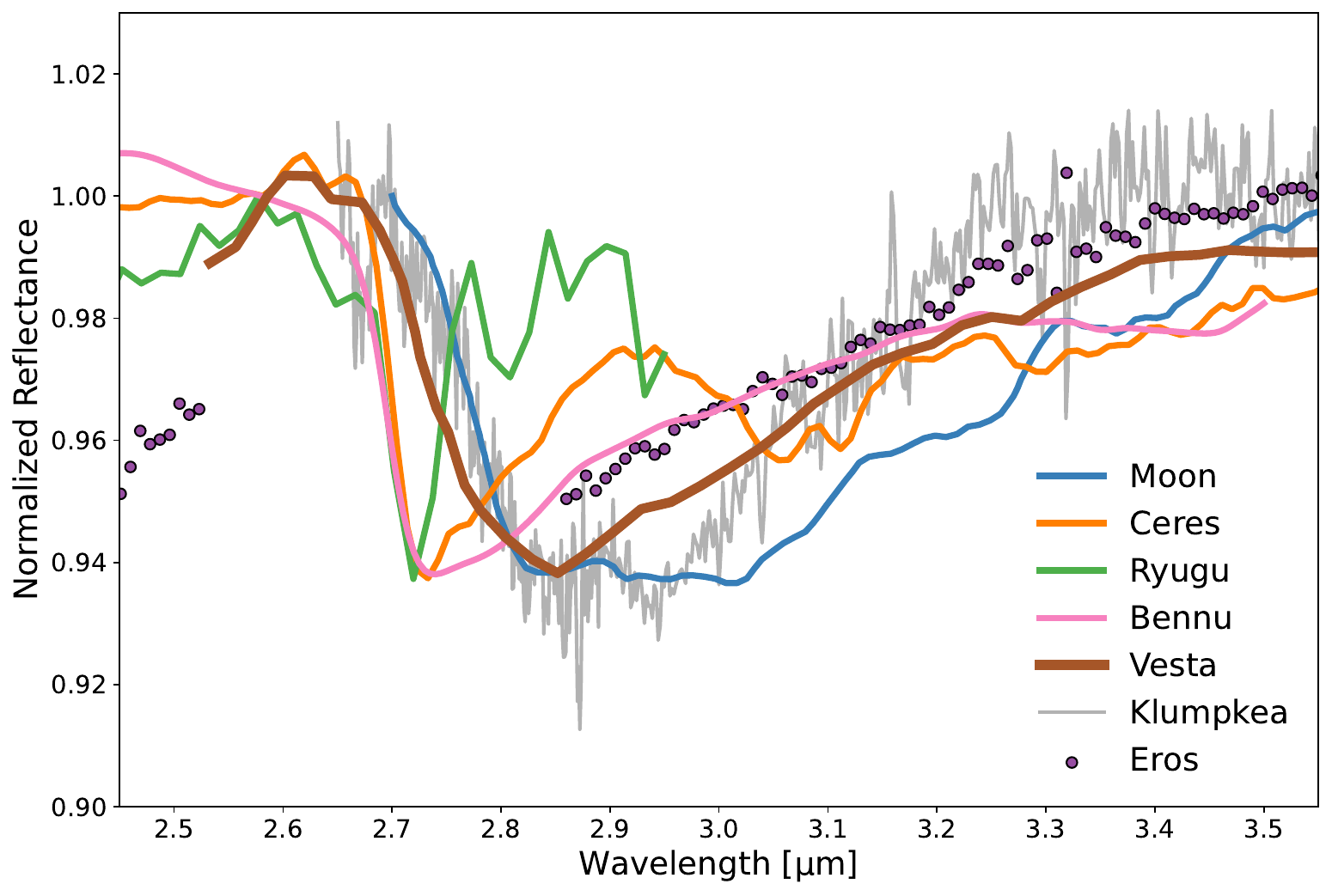}
    \caption{Comparison of the 2.85-$\mu$m features of Henan (top) and Klumpkea (bottom) from JWST with similar features seen in other solar system bodies. The normalized reflectance spectra of Bennu, Ryugu, Ceres, Vesta, and the Moon have been diluted to match the band depth of the JWST spectra. The spectrum of 433 Eros has been offset and lacks coverage between 2.55 to 2.85-$\mu$m due to atmospheric absorption.}
    \label{fig:space_compar}
\end{figure}

\begin{figure}
    \centering
    \includegraphics[width=\columnwidth]{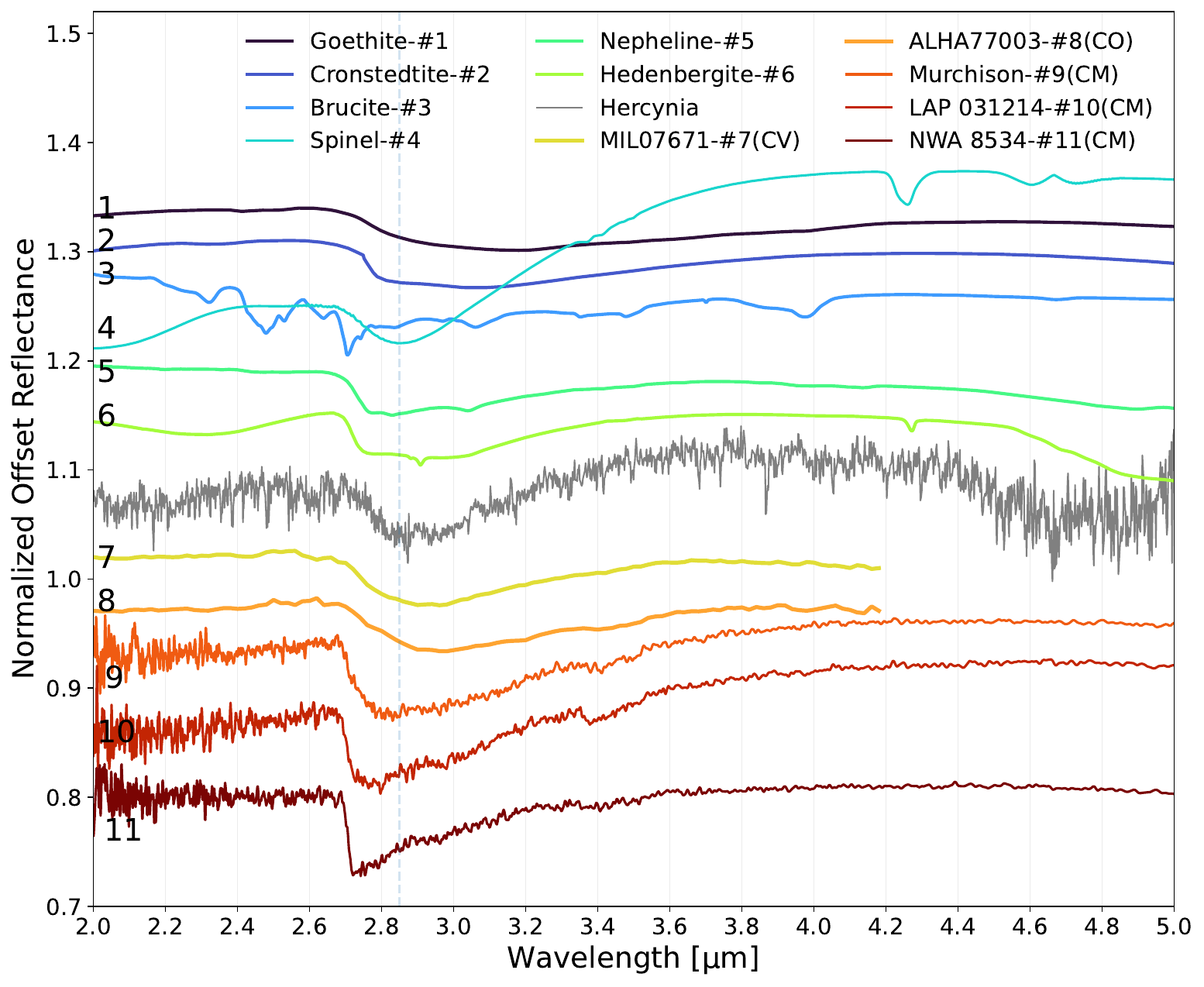}
    \includegraphics[width=\columnwidth]{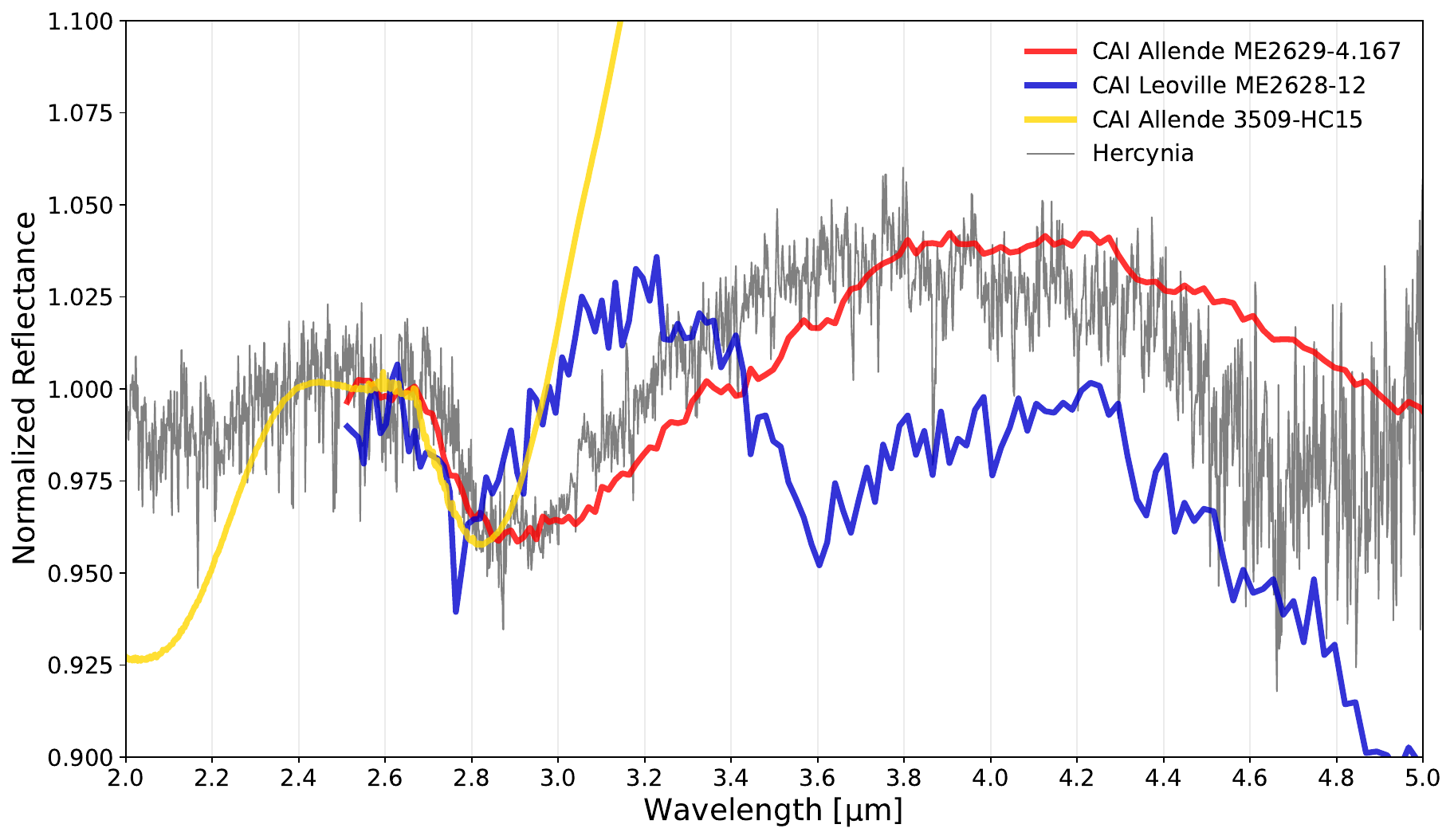}
    \caption{\textit{Top:} Comparison of the 2 to 5-$\mu$m spectrum of 458 Hercynia to the spectra of minerals and the spectra of aqueously altered CM, CO and CV meteorites. The particle sizes of all mineral samples are $<$ 45-$\mu$m except for goethite which is $<$ 125-$\mu$m. CM meteorites that have been heavily aqueously altered contain more phyllosillicates and exhibit a band center at shorter wavelengths. All meteorite samples were heated to minimize contamination from adsorbed terrestrial water. The meteorite and mineral spectra have been diluted to match the 3-$\mu$m band depth of 458 Hercynia. The dashed blue line corresponds to the band center of 458 Hercynia.
    \textit{Bottom:} Comparison of the 2 to 5-$\mu$m spectrum of 458 Hercynia to the spectra of three CAIs. The CAI spectra have also been diluted to match the 3-$\mu$m band depth of 458 Hercynia.}
    \label{fig:meteor}
\end{figure}

\section{Discussion}
\label{sec:disc}

In Figure \ref{fig:space_compar} we take the continuum-removed 3-$\mu$m features of 1040 Klumpkea and 2085 Henan as representative examples of our sample and compare them to those of small Solar System bodies and the Moon. Ceres, Ryugu, and Bennu are rich in carbonaeous material, associated with the C spectral type, and are spectrally dissimilar from L-types. The spectra of Ceres (obtained by the Dawn mission; \citealt{Russell2011}) shows a sharp asymmetric feature near $\sim$2.7-$\mu$m that is indicative of structural \ce{OH} in phyllosilicates \citep{deSanctis2015}. The spectrum of Ryugu (measured by Hayabusa2; \citealt{Tsuda2013}) shows a sharper asymmetric feature near $\sim$2.7-$\mu$m, which has been linked to \ce{OH}-bearing minerals \citep{Kitazato2019}. The Bennu spectrum, obtained with instruments onboard the OSIRIS-REx spacecraft \citep{Christensen2018, Reuter2018}, shows the same $\sim$2.7-$\mu$m phyllosilicate feature \citep{Hamilton2019, Simon2020a, Simon2020b}. We can see that the spectral features of our targets are largely in disagreement with the C-type spectral feature. The 3-$\mu$m band of our targets is centered at longer wavelengths, and it is broader than the band of the C-types. Although space weathering can shift the 2.7 $\mu$m feature to longer wavelengths \citep{Rubino2020}, the laboratory-observed shifts under irradiation are much smaller than the difference between the C- and L-type features.

The other asteroids in Figure \ref{fig:space_compar}, Vesta (V-type) and Eros (S-type), are stony, silicate-rich bodies inferred to be more closely related compositionally to the L-types. Vesta's spectrum, which was collected by the Dawn mission \citep{Russell2011}, exhibits a \ce{OH} feature that is interpreted to originate from low-velocity impacts of material containing hydrated mineral phases \citep{DeSanctis2012, McCord2012, Turrini2014, Massa2023}. Similarly, the spectrum of 433 Eros (obtained with the NASA Infrared Telescope Facility (IRTF); \citealt{Rivkin2018}) shows an absorption feature although it is obscured by Earth's atmosphere on the short-wavelength side. We can see that the 3-$\mu$m bands of these two stony asteroids more closely resemble the spectral feature of our targets. The band center of Vesta is consistent with those of our targets, although Vesta's absorption band begins at a shorter wavelength than our targets by nearly 0.05 $\mu$m. The 3-$\mu$m band center of Eros is unknown due to the lack of continuous coverage, but the slope of the feature on its long-wavelength side matches that of the features of our targets.

The final Solar System body in Figure \ref{fig:space_compar} is the Moon. The lunar spectrum, collected by the Deep Impact spacecraft during a flyby in 2009 \citep{Sunshine2009}, shows a hydration feature near 2.85-$\mu$m. This particular feature has been attributed to both \ce{OH} and \ce{H2O} with a solar wind origin \citep{Sunshine2009, Laferriere2022}. We can see that the spectral features of our targets are largely consistent with the lunar \ce{OH}/\ce{H2O} feature. The band centers of the Moon and the L-types match and the slope of the features at shorter wavelengths are the same. The width of the lunar feature and the width of band of the 2085 Henan feature are consistent, but the feature of 1040 Klumpkea is narrower on the long-wavelength side.

It is evident from our comparison that the 2.85-$\mu$m spectral features of our targets are most similar to the stony asteroids and the lunar \ce{OH}/\ce{H2O} features. The slight differences beyond 2.95-$\mu$m between the spectral features of the Moon and our targets may be due to differences in their \ce{H2O} content since \ce{H2O} exhibits a 3.0-$\mu$m absorption feature from both bending and stretching modes in the 2.8 to 3.1-$\mu$m range \citep[e.g.,][]{Bishop1994}. Furthermore, our comparison provides constraints on the source of the 2.85-$\mu$m \ce{OH} contribution. The absorption features of our targets are inconsistent with the $\sim$ 2.71 to 2.76-$\mu$m band centers of Bennu, Ryugu, and Ceres suggesting that hydrous mineral assemblages similar to those on C-class asteroids are likely not the source of the 2.85-$\mu$m component (Table \ref{tab:band_properties}, Figure \ref{fig:space_compar}). 

In Figure \ref{fig:meteor}, we further explore the potential causes of the 2.85-$\mu$m absorption by comparing laboratory spectra from meteorites and minerals to 458 Hercynia. Although previously we use 1040 Klumpkea and 2085 Henan for comparison, here we select 458 Hercynia because it is the intermediate case of these two end-members and it has a high signal-to-noise ratio, especially beyond 4.5-$\mu$m. The CM meteorite spectra \citep{Bates2020} span a sequence of objects that have been less aqueously altered (e.g., Murchison) to objects that have been heavily aqueously altered (e.g., NWA 8534). The \ce{OH} band center shifts from 2.78 to 2.82-$\mu$m in less altered carbonaceous chondrites with dominantly \ce{Fe} phyllosilicates to 2.71 to 2.73-$\mu$m for more altered meteorites with a greater proportion of \ce{Mg} phyllosilicates \citep{Beck2010, Takir2013, Bates2020}. But like C-type asteroid spectra, CM meteorites are inconsistent with our targets. Thus, it is unlikely the 2.85-$\mu$m absorption in our L-type spectra originates from phyllosilicates. The CO and CV meteorites \citep{Eschrig2021}, which have been linked to the L-types, also exhibit a hydration feature at longer wavelengths ($\sim$ 2.9-$\mu$m). The CV/CO feature is attributed to a convolution of \ce{OH} and \ce{H2O} \citep{Eschrig2021}, perhaps from non-phyllosilicates, but the mineralogical origin of the CV/CO feature is unclear.

In Figure \ref{fig:meteor} we compare laboratory spectra of phases found in CAIs ---  synthesized spinel \citep{Jackson2014}, nepheline and hedenbergite (both from RELAB spectral library; natural nepheline NE-EAC-003 and synthetic hedenbergite DL-CMP-082-A; \citealt{Klima2008}) --- and the CAIs Allende ME2629-4.167, Leoville ME2628-12, and Allende CAI Slab 6 3509-HC15 (from RELAB spectral library; TM-TJM-004), to the spectrum of 458 Hercynia. We also include laboratory spectra of the the minerals cronstedtite \citep{Bates2020}, goethite and brucite (both from RELAB spectral library; natural goethite JB-JLB-H58-A and natural brucite JB-JLB-944-A), which have been proposed as possible sources of 3-$\mu$m absorption on other targets \citep{Rivkin2006,Milliken2009, Beck2011}. The mineral spinel has a grain size $<$45-$\mu$m and $\sim$0.1 wt$\%$ of \ce{FeO}. The Allende ME2629-4.167 sample is a fine-grained inclusion, which is rich in spinel, while the Leoville ME2628-12 sample is a coarse-grained type-A inclusion, which is rich in melilite (for more information on coarse-grained and fine-grained CAIs see \citealt{MacPherson2014}). Furthermore, Allende (and by consequence, its inclusions) is oxidized via aqueous alteration, but Leoville (and its inclusions) is reduced. Lastly, Allende CAI Slab 6 3509-HC15 is classified as a fluffy type A CAI, with the major minerals being melilite and spinel \citep{Sunshine2008}.

In Figure \ref{fig:meteor} we show that the synthesized spinel, Allende ME2629-4.167, Allende CAI Slab 6 3509-HC15 and \ce{Fe}-cronstedtite spectra have an asymmetrical 2.8-$\mu$m absorption feature that is a good match to the 458 Hercynia spectral feature. In contrast Leoville ME2628-12, goethite and brucite show absorption features at 2.7-$\mu$m, 3.1-$\mu$m, and 3.05-$\mu$m respectively, that are a poor match. The 2.8-$\mu$m absorption feature seen in spinel is attributed to an electronic transition of $\mathrm{Fe^{2+}_{IV}}$ \citep{Cloutis2004, Jackson2014}, although it could originate from OH (see Figure 3 of \citealt{Jackson2014}). On cronstedtite, the 2.8-$\mu$m spectral feature is due to \ce{Fe-OH} and the 3.06-$\mu$m spectral feature is due to \ce{H2O}. In the case of Allende ME2629-4.167 and Slab 6 3509-HC15, the source of the 2.8-$\mu$m absorption is somewhat less certain. Since the two inclusions are enriched in spinel and Allende has been aqueously altered, their 2.8-$\mu$m feature could perhaps be a combination of spinel and/or \ce{OH}/\ce{H2O} absorption. Lastly, the minerals nepheline and hedenbergite also show a 2.8-$\mu$m absorption feature, but it is a poor match to that of 458 Hercynia.

Overall, the JWST spectra of the spinel-rich asteroids in our sample reveal that they have a 3-$\mu$m feature which likely originates from a combination of \ce{OH}/\ce{H2O} and/or the mineral spinel on the surfaces of these objects. As mentioned in Section \ref{sec:intro}, in general the mineral spinel may have been synthesized during the formation of CAIs or during a magmatic event on a parent body. \citealt{Sunshine2008} argued against an igneous origin for spinel in L-type asteroids since aluminous spinel rather than chromite is most consistent with ground-based spectra of L-type asteroids. However, aluminous spinel is also present on the Moon, where it is linked to magmatic events \citep{Pieters2011}. 

The presence of a 3-$\mu$m feature has different implications for our targets in the case of absorption from \ce{OH}/\ce{H2O} or from the mineral spinel. Assuming the 3-$\mu$m feature is an \ce{Fe-OH} feature (rather than an Fe transition in spinel), it suggests that our targets were aqueously altered, were contaminated by the infall of hydrated material and/or have experienced solar wind implantation of H. If the spinel on their surfaces is in the form of CAIs, aqueous alteration is expected. Fe in CAIs is primarily seen in aqueously altered samples, with Fe being thought to be introduced by the fluids into the CAIs \citep{Krot1995} because it does not condense at the high-temperatures at which the primary CAI minerals form \citep{Grossman1972}. Furthermore, models for the early Solar System also show that high-temperature materials like CAIs in the disk midplane were transported outward \citep{ciesla2007}, where volatiles were condensing. During this period, radiogenic heat from $^{26}$Al would have triggered aqueous alteration, as proposed for parent bodies of CM/CI chondrites \citep[e.g.,][]{Grimm1989, Grimm1993}. Alternatively, if the spinel originated in magmatic processes within a large parent body, aqueous alteration would be unexpected, though trace volatiles might be incorporated into magmas. In either case, contamination from impacts or solar wind implantation are plausible candidates for the feature on the L-types and potentially the S-types and other silicate bodies more broadly. Finally, it remains possible that the 2.85-$\mu$m feature is largely a spinel mineral feature with no OH present.

Additional laboratory work with spinel samples is needed to clarify whether the cause of the 2.85-$\mu$m absorption in spinel is due to an electronic transition or trace OH in a nominally anhydrous mineral sample. Additionally, future spectral modeling of the 3-$\mu$m feature jointly with the 1 and 2 $\mu$m features on the L-types objects could establish whether the abundances of spinel needed to match the three features are consistent, which would indicate whether the 3-$\mu$m feature largely originates from spinel or whether an additional contribution from \ce{OH}/\ce{H2O} or infall of material/solar wind implantation is required. Finally, a future 3-$\mu$m survey with JWST of stony S-type asteroids would help distinguish between origin scenarios by revealing whether the 2.85-$\mu$m feature seen in the L-types is unique to this type or general to silicate-rich bodies, as hinted at by the strong similarities between the 3-$\mu$m features in our targets and the Moon, Eros, and Vesta. Either case would have important implications for the availability of volatiles at objects that formed in the inner Solar System. 

\section{Conclusion}
\label{sec:conc}

In this work, we explored the surface composition of five spinel-rich asteroids. We present 1.7 to 5.1-$\mu$m spectra with JWST NIRSpec at a spectral resolution of R$\sim$1000. Our observations reveal a 2-$\mu$m absorption feature in some of these objects, which is consistent with previous ground-based observations. The 2-$\mu$m absorption feature has been attributed to spinel, which could be in the form of CAIs on the surfaces of these objects. Furthermore, JWST's improved sensitivity and continuous wavelength coverage have enabled the detection of a $\sim$ 2.85-$\mu$m absorption feature in all five targets. The 2.85-$\mu$m absorption feature is distinct from the 2.7-$\mu$m feature typically seen on C type asteroids and associated with aqueous alteration. Instead, the 2.85-$\mu$m feature generally resembles the absorption features seen on the Moon and Vesta, which are indicative of \ce{OH} and/or \ce{H2O}. However, given the evidence for the existence of spinel on these objects, it may be the case that the 2.85-$\mu$m absorption feature is instead caused by an electronic transition of Fe in spinel. These two cases have different implications for the origin of the mineral spinel and more broadly the geologic histories of these bodies. Future studies on deciphering the exact species responsible for the 2.85-$\mu$m absorption feature are thus critical.

\section*{Acknowledgements}
We thank the anonymous referee for a helpful report. We also thank Jessica Sunshine and Joseph Masiero for helpful discussions. The authors acknowledge support from the Caltech Center for Comparative Planetary Evolution (3CPE). This work is based on observations made with the NASA/ESA/CSA James Webb Space Telescope. The JWST data were obtained from the Mikulski Archive for Space Telescopes at the Space Telescope Science Institute, which is operated by the Association of Universities for Research in Astronomy, Inc., under NASA contract NAS 5-03127 for JWST. These observations are associated with program GO 2361 and can be accessed via \dataset[DOI: 10.17909/vjwg-7z31]{https://doi.org/10.17909/vjwg-7z31}. Support for program GO 2361 was provided by NASA through a grant from the Space Telescope Science Institute, which is operated by the Association of Universities for Research in Astronomy, Inc., under NASA contract NAS 5-03127. FLHT is grateful for support from a Packard Fellowship. CAI samples for this study were generously provided by P. Heck and J. Holstein at the Robert A. Pritzker Center for Meteoritics and Polar Studies at the Field Museum of Natural History in Chicago.

\newpage
\bibliography{Ltypes}{}
\bibliographystyle{aasjournal}

\end{document}